# INTERDISCIPLINARY INTEGRATION OF REMOTE SENSING — A REVIEW WITH FOUR EXAMPLES

*Zichen Jin*

School of Aeronautics and Astronautics, University of Electronic Science and Technology of China

**ABSTRACT**

As a high-level discipline, the development of remote sensing depends on the contribution of many other basic and applied disciplines and technologies. For example, due to the close relationship between remote sensing and photogrammetry, remote sensing would inevitably integrate disciplines such as optics and color science. Also, remote sensing integrates the knowledge of electronics in the conversion from optical signals to electrical signals via CCD (Charge-Coupled Device) or other image sensors. Moreover, when conducting object identification and classification with remote sensing data, mathematical morphology and other digital image processing technologies are used. These examples are only the tip of the iceberg of interdisciplinary integration of remote sensing. This work briefly reviews the interdisciplinary integration of remote sensing with four examples — ecology, mathematical morphology, machine learning, and electronics.

## 1. INTRODUCTION

As a high-level discipline, the development of remote sensing depends on the contribution of many other basic and applied disciplines and technologies. For example, due to the close relationship between remote sensing and photogrammetry, remote sensing would inevitably integrate disciplines such as optics and color science. Also, remote sensing integrates the knowledge of electronics in the conversion from optical signals to electrical signals via CCD (Charge-Coupled Device) or other image sensors. When conducting object identification and classification with remote sensing data, mathematical morphology and other digital image processing technologies are used. In the further analysis of remote sensing data, the knowledge of high-performance computing, machine learning, pattern recognition, and other high technologies in the field of data science plays a vital role. If remote sensing data is utilized to analyze the status of organisms such as crops, then the integration of remote sensing with relevant knowledge in life sciences is needed. Moreover, the transmission of remote sensing satellites' data with the Earth requires the use of radio frequency telecommunication technology.

These examples are only the tip of the iceberg of interdisciplinary integration of remote sensing. Therefore, remote sensing is certainly a science as well as a technology of multidisciplinary integration. This work briefly reviews the interdisciplinary integration of remote sensing via four disciplines as concrete examples — ecology, mathematical morphology, machine learning, and electronics.

## 2. INTERDISCIPLINARY INTEGRATION OF REMOTE SENSING WITH ECOLOGY

### 2.1. Remote sensing for forest and vegetation biomass estimation

Forest biomass is closely related to the carbon sources and sinks of forest ecosystems. Accurate estimation of forest biomass in large regions is of great significance for studying the carbon cycle of terrestrial ecosystems. Remote sensing is macroscopic, integrated, dynamic, rapid, and repeatable, and its waveband information and forest biomass structure have certain correlation. Remote sensing has become the main method for regional forest biomass estimation. [1]–[3]

*2.1.1. Overview of global development*
Optical remote sensing uses the differences in reflectance spectra of different types of plants or the same type of plants in different growth stages to establish biomass estimation models. Typical satellites include Landsat7, Terra, and Aqua from the US, and CartoSat1 from India [2].

Light Detection and Ranging (LiDAR) emits laser pulses that interact with the ground features, and the height of trees can be calculated based on the echo information. A model of the relationship between biomass and tree height of various forest trees has been established based on ground survey data, and the biomass can be obtained via this model. The GLAS (Geoscience Laser Altimeter System) on ICESat (Ice, Cloud, and Land Elevation Satellite) has been widely used for forest biomass estimation. The Advanced Topographic Laser Altimeter System on ICESat2 and the Global Ecosystems Dynamics Investigation LiDAR system on the International Space Station have improved the spot size and density compared to GLAS, providing support for obtaining high-resolution forest structure information [4]–[6].

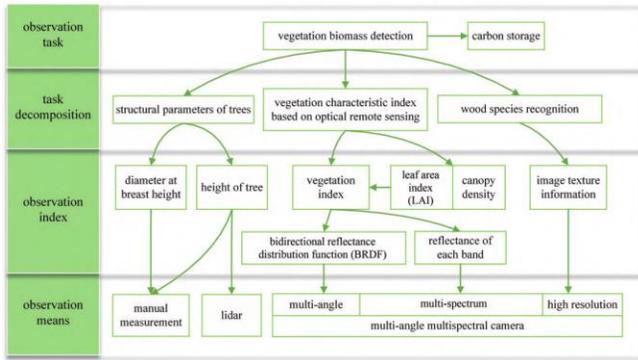

Fig. 1. Schematic diagram of vegetation biomass observation means.

Synthetic Aperture Radar (SAR) uses a microwave band which can penetrate the forest canopy and interact with tree trunks and branches to obtain the backscattered electromagnetic wave signals. Typical representatives are the ENVISAT (Environmental Satellite) and the Advanced Land Observing Satellite 2 [5], [7].

*2.1.2. Essential metrics to detect*
Remote sensing of forest biomass is mainly oriented to the applications of ecological construction and management. On the one hand, attention needs to be paid to the carbon stock of forest ecosystem and the current state of forest resources. On the other hand, the stability, energy balance, nutrient cycle, and productivity of the ecosystem, which can be shown by the variations in biomass volume, need to be focused on. In order to achieve high-precision remote sensing, it is also necessary to correct the influence of atmospheric environment. Therefore, the main detection metrics of forest biomass remote sensing are [2]:

(1) Volume of existing biomass.

(2) Variations in the volume of biomass: Periodic observations of ecosystems can be conducted to monitor and estimate changes. The amount of variation in forest biomass is mainly caused by photosynthesis. During the photosynthetic reaction of plants, there is a partial loss of absorbed light energy, which is emitted by chlorophyll molecules as a longer wavelength light signal called Solar-Induced chlorophyll Fluorescence (SIF). The intensity of SIF can characterize the intensity of vegetation photosynthesis and thus reflect the amount of biomass change [8].

(3) Auxiliary parameters: For example, via the observation of atmospheric aerosol, parameters such as the optical thickness of atmospheric aerosol can be obtained to correct the spectral curve of the reflectance of features, thus improving the accuracy of forest detection.

**2.2. Remote sensing for biodiversity monitoring**

Initial efforts to understand biodiversity have largely focused on the exploration of animal and plant species by taxonomists. Remote sensing measures the energy that is reflected and emitted from the Earth's surface. The spectral variation hypothesis is related to the use of particular sensors providing high spatial resolution images and is able to measure different signals about the phenology, biochemistry and structure of vegetation to get information at species level [9], [10]. Remote sensing is the only state of the art technology, able to provide global coverage and continuous measures about the condition of biodiversity. Remote sensing is increasingly supplemented by *in situ* sensing with cameras on stationary objects, Unmanned Aerial Vehicles (UAVs), smart phones, and electronic transmission tags [11]. Buhne and Pettorelli (2018) [12] reviewed multispectral and radar remote sensing data fusion in biodiversity monitoring. Kattenborn *et al.* [13] has assessed the potential of UAV for data acquisition on species cover of woody invasive species and upscaled the estimated species cover to the spatial scale of Sentinel-1 and Sentinel-2. The study of Duporge *et al.* [14] applied a neural network to automatically detect and count the African elephants using WorldView-3 and 4 data.

**2.3. Summary of usages of remote sensing in the fields of Ecology, Biodiversity, and Conservation (EBC)**

In general, ecological research refers to the investigation of organisms and their surrounding environment, including biotic and abiotic entities [15]. Biodiversity should be related to not only the variation of life forms, but also the ecological complexes of which they are a part. Conservation has become an indispensable way of dealing with ecosystem degradation, which have a significantly negative effect on biodiversity [16], [17]. Remote sensing, the science of obtaining information via noncontact recording, has swept the fields of EBC. Remote sensing can provide consistent long-term Earth observation data at scales from the local to the global domain. In addition, remote sensing is not labor-intensive and time-consuming, compared with field-based observations. The review papers of Kerr and Ostrovsky [18] and Turner *et al.* [19] have been cited thousands of times by scientists from around the world who are involved in remote sensing of EBC. Turner *et al.* stated two categories of approaches, namely direct and indirect remote sensing approaches.

## 3. INTERDISCIPLINARY INTEGRATION OF REMOTE SENSING WITH MATHEMATICAL MORPHOLOGY

**3.1. Introduction to mathematical morphology**

Mathematical morphology has been fully developed on binary images [20]. Mathematical morphology operators (erosion ($-$), dilation ($+$), opening ($\oplus$), and closing ($\otimes$)) act at a local level. These operators are defined between the input image ($D$) and a structuring element ($S$). Erosion, denoted "$E = D - S$", is the most important operator, because other morphological operators can be easily derived from it by using the complement, $D^C$, of the image [21]:

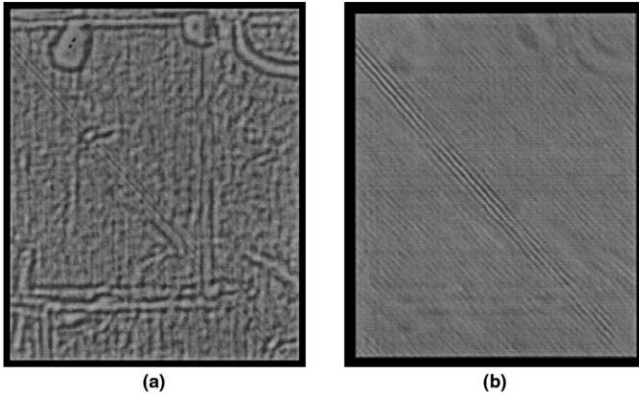

Fig. 2. The erosion based on the "correlation" using two structuring elements, (a) tower and (b) power line.

$$D + S = (D^C - S)^C,$$
$$D \oplus S = (D - S) + S,$$
$$D \otimes S = (D + S) - S.$$

The erosion is obtained by computing the local correlation between the structuring element *S* and all sub-images of *D*; this operation can be considered as shape matching [22] and *S* carries the shape to be retrieved. *S* can be computed from a training set of representative input data, or it can be built in a synthetic way. In the example shown in Fig. 2, synthetic models of towers and power lines are used. In the case of erosion of binary images, the image values take value "1" if there is a perfect agreement between the structuring element and the image, "0" otherwise.

Several methods have been proposed to extend mathematical morphology to gray level images. One approach considers gray level images as three dimensional (3D) objects (the third dimension is the intensities in each pixel), the structuring element is also a 3D image and the erosion is defined as before [22]. The main idea is of enhancing values of *D*, according to a measure of similarity between *D* and *S*. For example, the extension of standard binary operators is achieved using min-max operators [23]. In 1993, an approach based on fuzzy operators is proposed [24], where the image is transformed in a fuzzy set, so that its values can be interpreted as grade of membership in the set of high-valued pixels.

### 3.2. Application of mathematical morphology for remote sensing images

The unique role of mathematical morphology in the quantitative description and analysis of geometric features has made its research in remote sensing image processing quite deep, and the main work accomplished includes: morphological pattern recognition methods, morphological classification methods, morphological sampling and interpolation, morphological sequence decomposition, dynamic edge detection, morphological feature extraction, and fast morphological transformation methods [23].

For visible light and infrared data, directional morphological filtering can be used to extract the geometric structure of roads, rivers and towns [25]. The area of building clusters in remote sensing images can be extracted by resampling morphological gradients, and the morphological open and closed operations are good for eliminating image noise and smoothing image edges. Sequential morphological filtering can be used to extract the area of vortex range in infrared ocean images. For SAR images, using sequential morphological direction open and closed filtering can eliminate the speckle noise and extract the road network in the images. For digital elevation model, the data can be regarded as grayscale images, and the morphological operators are suitable for grayscale image processing.

### 3.3. Application example: road extraction

With the application of high-resolution remote sensing images, more information can be extracted from remote sensing images, but the large amount of content in high-resolution images makes information extraction more difficult [26], [27]. Extracting roads from remote sensing images is one of the problems of current research. Kahraman *et al.* [28] reviewed road detection methods and evaluated their advantages and disadvantages. Dai Jiguang *et al.* [29]–[31] classified the road detection methods into five categories: template matching-based methods, knowledge-based methods, object-oriented methods, deep learning methods, and mathematical morphology methods.

Since the 1980s, many scholars have gradually used mathematical morphology methods for processing remote sensing images, and since then, many methods for extracting roads in remote sensing images using mathematical morphology have been proposed. Valero *et al.* [25], [32] proposed a method for road extraction from high-resolution remote sensing images based on advanced oriented morphological operators, which extracts linear geometric pixel information by constructing path opening and closing operators to classify each pixel as road or non-road, and these operators do not depend on the choice of structural element shapes. Ma Ronggui *et al.* [33] proposed an automatic road extraction method for fuzzy aerial images, which first uses a multi-scale Retinex algorithm to enhance high-resolution low-contrast images, then segments the enhanced images using a modified Canny edge detection operator, and finally adjusts straight and curved road breaks using Hough linear transform and morphological operators.

## 4. INTERDISCIPLINARY INTEGRATION OF REMOTE SENSING WITH MACHINE LEARNING

### 4.1. Remote sensing image classification and recognition method

Remote sensing image classification and recognition is to use computer to analyze the spectral information and spatial information of various features in remote sensing image, and through feature screening, classify the image elements in the image into different categories according to certain rules, and then carry out information marking of real scenes. The common classification methods are supervised classification method and unsupervised classification method.

*4.1.1. Supervised classification*
Supervised classification requires prior knowledge of features and labels, and the trained model will record the features of remote sensing images and make classification prediction based on these features to achieve specific classification of images [14]. Common supervised classification methods include Support Vector Machine (SVM) and neural network classification methods.

*4.1.2. Unsupervised classification*
Unsupervised classification requires only prior knowledge of features and no labeling information, and it takes clustering as the basic idea and constructs a division based on association rules. One typical algorithm is the K-Means algorithm, which is divided into four steps: the first step is to determine the K initial cluster centers; the second step is to group the K samples closest to the center in a certain class; the third step is to recalculate the cluster centers of the class; the fourth step is to repeat the process of steps 1 to 3 until convergence, *i.e.*, the cluster centers no longer change. Another typical unsupervised learning algorithm is the ISODATA algorithm, which adds a class merging and splitting mechanism and has a more complex structure.

**4.2. Remote sensing image classification and recognition with Convolutional Neural Network (CNN)**

Chen Wenkang (2016) [34] trained and tested rural building and non-building images under the CaffeNet learning framework, and the recognition rate reached 95%. In 2017, Zhao Mandan *et al.* [35] analyzed the pixel-by-pixel spectral information by building a 5-layer neural network, and then provided a full-spectrum dataset at the input side and introduced a cost function to complete the feature extraction and classification of spectral information, with a classification accuracy of 90.16%; Luo Jianhua [36] used the spatial neighborhood structure information of all pixel points as the input of CNN model and designed the activation function ReLU, and proved through experiments that the mini-batch stochastic gradient descent method can improve the CNN classification accuracy, and the classification accuracy reached 97.57%; Du Jing [37] established a watershed identification model using Deep Convolutional Neural Network (DCNN), analyzed the UAV high-resolution remote sensing images by using MSER algorithm, and imported the DCNN water body identification model by targeting the target area to be identified, acquiring an identification accuracy as high as 95.36%. Zhu Yuanjie *et al.* (2020) [38] built a CNN model to train images based on specific situational semantics for the characteristics of Jianye district (in Nanjing, Jiangsu province, China), such as complicated urban green space categories and obvious regional differences, so as to achieve automatic classification of various types of green spaces, and the classification accuracy was proved to be 87.74% through experiments. Wang Jia'nan *et al.* (2021) [39] proposed an optical remote sensing image classification method based on the dual branching structure of visual converter and graph convolution network. The method firstly chunked the image, and then used position coding and visual converter to encode the features of the image; meanwhile, the hyperpixel segmentation was performed on the remote sensing image, and the features of the convolutional neural network corresponding to each hyperpixel were pooled and used as nodes in the graph structure, and the graph convolutional network was used to model the internal graph structure of the scene; finally, the features generated from the two branches were fused to form the final features and used for classification. And the effectiveness of the proposed method in remote sensing scene classification is verified by comparison experiments. Xu Shanshan *et al.* (2022) [40] selected a DCNN model to detect vegetation areas in high-resolution remote sensing images, firstly analyzing different optimizers and conducting comparison experiments by setting different convolutional kernel sizes; then investigating the number of network layers; and finally performing vegetation area detection with the constructed DCNN model.

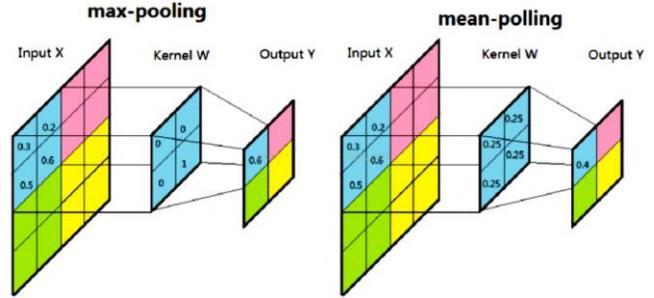

Fig. 3. Schematic diagram of CNN architecture.

**5. INTERDISCIPLINARY INTEGRATION OF REMOTE SENSING WITH ELECTRONICS**

**5.1. Introduction to satellite borne integrated electronic system for remote sensing satellites**

According to recent research progresses, satellite borne integrated electronic system can be defined from the following aspects [41].
(1) Design and purpose: The satellite borne integrated electronic system is an integrated system designed by system engineering method and developed in a unified modular environment, which coordinates and controls various sensors, actuators and other resources of spacecraft platforms and payloads to work in an orderly manner, provides information processing and services for the whole spacecraft, and manages the whole spacecraft mission operation and safety things in a unified manner [42], [43].

(2) Core: The core of a satellite borne integrated electronic system is a trinity service support framework of electricity, information, and control, providing a full range of services for the satellite platform and payload. Hardware and software of each subsystem still exist, but information and management form a unified manner [44], [45].

(3) Functions: The functions of satellite borne integrated electronic systems includes: mission operation (telemetry and remote control, attitude and orbit control, communication, *etc.* [45]), payload management (payload management, mission planning, *etc.*), satellite service management (thermal control management, information service, satellite time service, *etc.*), resource management (energy management, storage management, computing resource management, *etc.*), and safety management (status monitoring, fault diagnosis, fault reorganization, task degradation, *etc.*).

(4) Features: The essential features of satellite borne integrated electronic systems are: (a) unification — unified design of the entire electronic system in a common modular environment; (b) optimization — optimization of the entire electronic system through top-level design; (c) standardization — unified standards in system architecture, protocols, interfaces, *etc.*, to achieve standardization of components, products, services and systems; and (d) efficiency — achieving flexibility and reusability of the system and its equipment via the decomposition of the system's functions [41].

(5) Developing method: In a satellite borne integrated electronic system, all the components are placed in a complete and reasonable architecture, using a top-down system engineering approach to complete the development of the system; the most critical aspects of the development process of the satellite borne integrated electronic system are: requirements analysis, system function definition, system modeling, detailed design, design verification and optimization, hardware and software development, system integration and testing [42], [43].

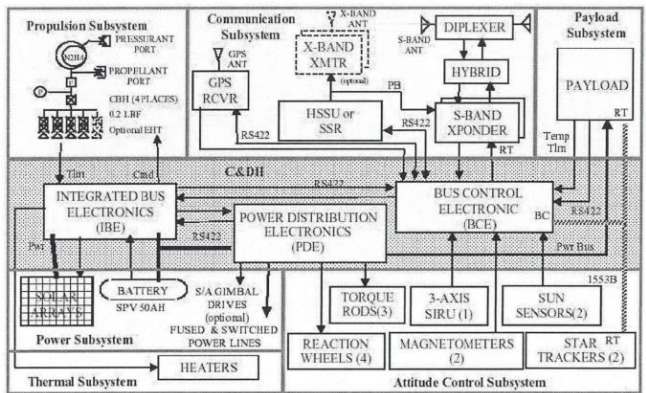

Fig. 4. An example of integrated electronic system on remote sensing satellite — the LM-900 platform.

## 5.2. Prospects and inspirations for future development

By comparing and analyzing the development of satellite borne integrated electronic systems, as well as the current space mission requirements for satellite design, the future development trends of satellite borne integrated electronic systems are as follows.

(1) Network system develops towards standardized protocols, buses, and interfaces. The widespread application of standard information processing and communication protocol systems such as the Consultative Committee for Space Data Systems (CCSDS) will play an important role in intra-planetary, terrestrial, and interplanetary information transfer and spacecraft information fusion for spacecraft in orbit. In order to enable systems with different buses and interfaces to interact with each other, CCSDS has defined a reference model for the on-board interface through the study of Spacecraft Onboard Interface Service (SOIS), divided the services of each standard layer, and standardized the protocol and interface design [46].

(2) Satellite borne hardware develops towards standardization, modularization, integration, and miniaturization. Unlike the traditional satellite electronic system, the new generation of integrated electronic system adopts the modular design idea according to function, the whole system consists of standardized common modules, which are divided according to their functions. The modular design is adopted in the hardware architecture, using standard boards, standard chassis and standard internal buses to improve the integration of the system, so that the new generation of integrated electronic system with good openness, can better achieve functional reduction and expansion, improve the flexibility of the use of integrated electronic system, the platform has good scalability and compatibility, so that it can adapt to the needs of the expansion of satellite functions [42], [44].

(3) Integrated electronics, on top of being in the management of the daily operation of the satellite, also takes more responsibilities in mission control. The future integrated electronic system will integrate its original functions such as measurement and control, attitude and orbit control, thermal control [43], power management, payload management, while more attention will be paid to the planning and management of satellite missions. The spacecraft's information, resources, operation, and crisis managements and other autonomous management functions will reduce the burden of long-term in-orbit management and effectively improve the control capability of the satellite.

## 6. REFERENCES

[1] N. B. Mishra, K. A. Crews, and G. S. Okin, "Relating spatial patterns of fractional land cover to savanna vegetation morphology using multi-scale remote sensing in the central kalahari," Int. J. Remote Sens., vol. 35, no. 6, pp. 2082–2104, Mar. 2014, doi: 10.1080/01431161.2014.885666.
[2] Li Bin and Liu Kening, "Forest Biomass Estimation Based on

UAV Optical Remote Sensing," FOREST ENGINEERING, vol. 38, no. 5, pp. 83–92, 2022, doi: 10.3969/j.issn.1006-8023.2022.05.011.
[3] Cao Haiyi, Qiu Xinyi, and He Tao, "Review on Development of Forest Biomass Remote Sensing Satellites," Acta Optica Sinica, vol. 42, no. 17, p. 1728001, 2022, doi: 10.3788/AOS202242.1728001.
[4] A. M. Lechner, G. M. Foody, and D. S. Boyd, "Applications in remote sensing to forest ecology and management," One Earth, vol. 2, no. 5, pp. 405–412, May 2020, doi: 10.1016/j.oneear.2020.05.001.
[5] E. Chraibi, H. Arnold, S. Luque, A. Deacon, A. E. Magurran, and J.-B. Feret, "A remote sensing approach to understanding patterns of secondary succession in tropical forest," Remote Sensing, vol. 13, no. 11, Jun. 2021, doi: 10.3390/rs13112148.
[6] U. Panne, "Laser remote sensing," Trac-Trends in Analytical Chemistry, vol. 17, no. 8–9, pp. 491–500, Oct. 1998, doi: 10.1016/s0165-9936(98)00054-5.
[7] N. B. Mishra and K. A. Crews, "Mapping vegetation morphology types in a dry savanna ecosystem: Integrating hierarchical object-based image analysis with random forest," International Journal of Remote Sensing, vol. 35, no. 3, pp. 1175–1198, Feb. 2014, doi: 10.1080/01431161.2013.876120.
[8] K. Kohzuma, K. Sonoike, and K. Hikosaka, "Imaging, screening and remote sensing of photosynthetic activity and stress responses," J. Plant Res., vol. 134, no. 4, pp. 649–651, Jul. 2021, doi: 10.1007/s10265-021-01324-1.
[9] Tian J.-Y. et al., "Application of spectral diversity in plant diversity monitoring and assessment," Chinese Journal of Plant Ecology, vol. 46, no. 10, pp. 1129–1150, 2022, doi: 10.17521/cjpe.2022.0077.
[10] Zhang Xiaobo, Guo Lanping, Huang Luqi, Zhu Shoudong, and Ma Weifeng, "Application of hyperspectral remote sensing in field of medicinal plants monitoring research," China Journal of Chinese Materia Medica, vol. 38, no. 9, pp. 1280–1284, 2013, doi: 10.4268/cjcmm20130903.
[11] He Cheng, Feng Zhongke, Yuan Jinjun, Wang Jia, Gong Yinxi, and Dong Zhihai, "Advances in the Research on Hyperspectral Remote Sensing in Biodiversity and Conservation," Spectroscopy and Spectral Analysis, vol. 32, no. 6, pp. 1628–1632, 2012, doi: 10.3964/j.issn.1000-0593(2012)06-1628-05.
[12] H. Schulte To Bühne and N. Pettorelli, "Better together: Integrating and fusing multispectral and radar satellite imagery to inform biodiversity monitoring, ecological research and conservation science," Methods Ecol Evol, vol. 9, no. 4, pp. 849–865, Apr. 2018, doi: 10.1111/2041-210X.12942.
[13] T. Kattenborn, J. Lopatin, M. Förster, A. C. Braun, and F. E. Fassnacht, "UAV data as alternative to field sampling to map woody invasive species based on combined sentinel-1 and sentinel-2 data," Remote Sensing of Environment, vol. 227, pp. 61–73, Jun. 2019, doi: 10.1016/j.rse.2019.03.025.
[14] I. Duporge, O. Isupova, S. Reece, D. W. Macdonald, and T. Wang, "Using very-high-resolution satellite imagery and deep learning to detect and count african elephants in heterogeneous landscapes," Remote Sens Ecol Conserv, vol. 7, no. 3, pp. 369–381, Sep. 2021, doi: 10.1002/rse2.195.
[15] K. Wang, S. E. Franklin, X. Guo, and M. Cattet, "Remote sensing of ecology, biodiversity and conservation: A review from the perspective of remote sensing specialists," Sens., vol. 10, no. 11, pp. 9647–9667, Nov. 2010, doi: 10.3390/s101109647.
[16] P. J. Stephenson et al., "Priorities for big biodiversity data," Frontiers in Ecology and the Environment, vol. 15, no. 3, pp. 124–125, Apr. 2017, doi: 10.1002/fee.1473.
[17] A. K. Skidmore et al., "Priority list of biodiversity metrics to observe from space," Nature Ecology & Evolution, vol. 5, no. 7, pp. 896–906, May 2021, doi: 10.1038/s41559-021-01451-x.
[18] J. T. Kerr and M. Ostrovsky, "From space to species: Ecological applications for remote sensing," Trends in Ecology & Evolution, vol. 18, no. 6, pp. 299–305, Jun. 2003, doi: 10.1016/S0169-5347(03)00071-5.
[19] W. Turner, S. Spector, N. Gardiner, M. Fladeland, E. Sterling, and M. Steininger, "Remote sensing for biodiversity science and conservation," Trends Ecol. Evol., vol. 18, no. 6, pp. 306–314, Jun. 2003, doi: 10.1016/s0169-5347(03)00070-3.
[20] J. Serra, Image analysis and mathematical morphology, vol. 1. London: Academic Press Inc., 1982.
[21] F. Li, H. Wang, W. Jia, and Z. Liu, "Remote sensing image segmentation of ulan buh desert based on mathematical morphology," presented at the International Conference on Computational Materials Science (CMS 2011), in Advanced Materials Research, vol. 268–270. Apr. 2011, pp. 1332–1338. doi: 10.4028/www.scientific.net/AMR.268-270.1332.
[22] F. Yeong-Chyang Shih and O. R. Mitchell, "Decomposition of gray-scale morphological structuring elements," Pattern Recognition, vol. 24, no. 3, pp. 195–203, Jan. 1991, doi: 10.1016/0031-3203(91)90061-9.
[23] R. M. Haralick, S. R. Sternberg, and X. Zhuang, "Image analysis using mathematical morphology," IEEE Trans. Pattern Anal. Mach. Intell., vol. PAMI-9, no. 4, pp. 532–550, Jul. 1987, doi: 10.1109/TPAMI.1987.4767941.
[24] V. Di Gesú, "Integrated fuzzy clustering," Fuzzy Sets and Systems, vol. 68, no. 3, pp. 293–308, Dec. 1994, doi: 10.1016/0165-0114(94)90185-6.
[25] S. Valero, J. Chanussot, J. A. Benediktsson, H. Talbot, and B. Waske, "Directional mathematical morphology for the detection of the road network in very high resolution remote sensing images," in 2009 16th IEEE International Conference on Image Processing (ICIP), Cairo, Egypt: IEEE, Nov. 2009, pp. 3725–3728. doi: 10.1109/ICIP.2009.5414344.
[26] Chen Chao, Qin Qiming, Chen Li, Wang Jinliang, Liu Mingchao, and Wen Qi, "Extraction of Bridge over Water from High-Resolution Remote Sensing Images Based on Spectral Characteristics of Ground Objects," Spectroscopy and Spectral Analysis, vol. 33, no. 3, pp. 718–722, 2013.
[27] Y. Yuksel, D. Maktav, and S. Kapdasli, "Application of remote-sensing technology to the relation of submarine pipelines and coastal morphology," Water Sci. Technol., vol. 32, no. 2, pp. 77–83, 1995, doi: 10.1016/0273-1223(95)00572-5.
[28] I. Kahraman, I. R. Karas, and A. E. Akay, "Road extraction techniques from remote sensing images: A review," Int. Arch. Photogramm. Remote Sens. Spatial Inf. Sci., vol. XLII-4/W9, pp. 339–342, Oct. 2018, doi: 10.5194/isprs-archives-XLII-4-W9-339-2018.
[29] Dai J. et al., "Development and prospect of road extraction method for optical remote sensing image," Journal of Remote Sensing (Chinese), vol. 24, no. 7, pp. 804–823, 2020, doi: 10.11834/jrs.20208360.
[30] J. Dai, T. Zhu, Y. Wang, R. Ma, and X. Fang, "Road extraction from high-resolution satellite images based on multiple descriptors," IEEE J. Sel. Top. Appl. Earth Observations Remote Sensing, vol. 13, pp. 227–240, 2020, doi: 10.1109/JSTARS.2019.2955277.
[31] J. Dai, T. Zhu, Y. Zhang, R. Ma, and W. Li, "Lane-level road extraction from high-resolution optical satellite images," Remote Sensing, vol. 11, no. 22, p. 2672, Nov. 2019, doi: 10.3390/rs11222672.
[32] S. Valero, J. Chanussot, J. A. Benediktsson, H. Talbot, and B. Waske, "Advanced directional mathematical morphology for the

detection of the road network in very high resolution remote sensing images," Pattern Recognition Letters, vol. 31, no. 10, pp. 1120–1127, Jul. 2010, doi: 10.1016/j.patrec.2009.12.018.

[33] M. Ronggui, W. Weixing, and L. Sheng, "Extracting roads based on retinex and improved canny operator with shape criteria in vague and unevenly illuminated aerial images," J. Appl. Remote Sens, vol. 6, no. 1, p. 063610, Dec. 2012, doi: 10.1117/1.JRS.6.063610.

[34] Chen Wenkang, "Remote Sensing Image Detection of Rural Buildings Based on Deep Learning Algorithm," Surveying and Mapping, ISSN: 1674-5019, vol. 39, no. 5, pp. 227–230, 2016.

[35] Zhao Mandan, Ren Zhiquan, Wu Gaochang, and Hao Xiangyang, "Convolutional Neural Networks for Hyperspectral Image Classification," Journal of Geomatics Science and Technology, vol. 34, no. 5, pp. 501–507, 2017, doi: 10.3969/j.issn.1673-6338.2017.05.013.

[36] Jianhua Luo, "The Application of Deep Learning in Dimensionality Reduction And Classification of Hyperspectral Image," Master, University of Electronic Science and Technology of China, 2018. [Online]. Available: https://kns.cnki.net/kcms2/article/abstract?v=2C6ioF1tvgXfqMzU-fBqdXJgADqcLTqjsdxAp2QbIgZ5KK47wV38jAOK31emFvVVtzZAgQFOPpY-fz01ShfHU1pF7cu8WXoMlgMavzb8qC7MNjJeb4DZHA==&uniplatform=NZKPT&language=gb

[37] Du Jing, "Deep Learning Based UAV Remote Sensing Image Water Body Identification," JIANGXI SCIENCE, vol. 35, no. 1, pp. 158-161+170, 2017, doi: 10.13990/j.issn1001-3679.2017.01.031.

[38] Zhu Yuanjie, Jiang Miaojun, Li Xiaotian, and Wang Xiaopo, "Greenfield Information Extraction for Scene Classification Based on Convolutional Neural Network," Beijing Surveying and Mapping, vol. 34, no. 12, pp. 1780–1784, 2020, doi: 10.19580/j.cnki.1007-3000.2020.12.025.

[39] Wang Jia'nan, Gao Yue, Shi Jun, and Liu Ziqi, "Scene Classification of Optical High-resolution Remote Sensing Images Using Vision Transformer and Graph Convolutional Network," ACTA PHOTONICA SINICA, vol. 50, no. 11, p. 1128002, 2021, doi: 10.3788/gzxb20215011.1128002.

[40] Xu Shanshan, Lyu Jingyan, and Chen Fangyuan, "Remote sensing vegetation detection method based on the deep convolutional neural network," Journal of Nanjing Forestry University (Natural Sciences Edition), vol. 46, no. 4, pp. 185–193, 2022, doi: 10.12302/j.issn.1000-2006.202009059.

[41] Zhenxing Liu, "Research on Integrated Avionics for Remote Sensing Satellite," Doctorate, University of Science and Technology of China, 2021. doi: 10.27517/d.cnki.gzkju.2021.000782.

[42] L. Jiang and P. Xu, "The simulation and verification platform for micro-satellite integrated electronic systems," in 2018 Eighth International Conference on Instrumentation & Measurement, Computer, Communication and Control (IMCCC), Harbin, China: IEEE, Jul. 2018, pp. 1434–1437. doi: 10.1109/IMCCC.2018.00296.

[43] G. Tanda, "An experimental study on the transient thermal response of an electronic equipment box for UAV remote sensing applications," J. Phys.: Conf. Ser., vol. 1599, no. 1, p. 012037, Aug. 2020, doi: 10.1088/1742-6596/1599/1/012037.

[44] L. Xing and B. W. Johnson, "Reliability Theory and Practice for Unmanned Aerial Vehicles," IEEE Internet Things J., vol. 10, no. 4, pp. 3548–3566, Feb. 2023, doi: 10.1109/JIOT.2022.3218491.

[45] Haibin Liu, Wenqiang Ji, and Junjie Hou, "A robust control method for load electronic system of autonomous remote sensing satellite in-orbit," in 2014 IEEE Conference and Expo Transportation Electrification Asia-Pacific (ITEC Asia-Pacific), Beijing, China: IEEE, Aug. 2014, pp. 1–5. doi: 10.1109/ITEC-AP.2014.6940802.

[46] F. Bartolini, "Watermarking techniques for electronic delivery of remote sensing images," Opt. Eng, vol. 41, no. 9, p. 2111, Sep. 2002, doi: 10.1117/1.1496787.